# Giant viruses in the oceans : the 4[th] Algal Virus Workshop

**Jean-Michel Claverie**

Structural & Genomic Information Laboratory, UPR 2589, IBSM, CNRS

31 chemin Joseph Aiguier, 13402 Marseille Cedex 20, and University of Mediterranee School of Medicine, 13385 Marseille Cedex 5, FRANCE

Tel: +33 4 91 16 45 48

Fax: +33 4 91 16 45 49

Jean-Michel.Claverie@igs.cnrs-mrs.fr

# Abstract

Giant double-stranded DNA viruses (such as record breaking *Acanthamoeba polyphaga Mimivirus*), with particle sizes of 0.2 to 0.6 μm, genomes of 300 kbp to 1.200 kbp, and commensurate complex gene contents, constitute an evolutionary mystery. They challenge the common vision of viruses, traditionally seen as highly streamlined genomes optimally fitted to the smallest possible –filterable- package. Such giant viruses are now discovered in increasing numbers through the systematic sampling of ocean waters as well as freshwater aquatic environments, where they play a significant role in controlling phyto- and bacterio-plankton populations. The 4$^{th}$ algal virus workshop showed that the study of these ecologically important viruses is now massively entering the genomic era, promising a better understanding of their diversity and, hopefully, some insights on their origin and the evolutionary forces that shaped their genomes.

The 4th Algal Virus Workshop (www.avw4.org) organized by Corina Brussaard and Herman Gons, and hosted by the Royal Netherlands Institute for Sea Research, was held in Amsterdam 17-21 april 2005. Though marine ecology rather than basic virology was the main focus of this meeting, exciting new results on the genomics of large/giant viruses kept turning up in many talks. In the context of a comparative study, Corina Brussaard (in collaboration with the US DoE) is herself sequencing a variety of *Micromonas pusilla* and *Phaeocystis globosa* dsDNA viruses some of them estimated to have a genome sizes up to **460 kb**.

In his overview, Curtis Suttle (University of British Columbia, Vancouver, Canada) pointed out that viruses (including RNA-, DNA-, prokaryotic and eukaryotic viruses) constitute a significant part of the biomass in ocean coastal waters (with up to 50 millions particles/ml, for a total estimate of 25 to 270 Megatons in the oceans) where they play a dominant role in the control of phyto- and bacterio-plankton populations, and hence on the production of oxygen and atmospheric dimethylsulphide, an important factor in climate regulation. Most of these viruses are uncharacterized [1].

Ironically, this is in a *freshwater* unicellular green alga that the best characterized large DNA virus *Paramecium bursaria chlorella* virus (PBCV-1), the prototype of the *Phycodnaviridae*, was isolated more than 20 years ago in Jim van Etten's laboratory (University of Nebraska, Lincoln)[2]. Liza Fitzgerald (Van Etten's laboratory) reported on the ongoing annotation of the genomic sequences of two new species of *Paramecium bursaria chlorella* viruses: NY-2A (infecting PBCV-1 host Chlorella species NC64A) and Chlorella Pbi virus MT325. NY-2A genome contains **368,683 bp**, making it the largest chlorella virus sequenced to date. Despite a 10% difference in size, the NY-2A genome and PBCV-1 genome (**330 kb**) exhibits a near perfect colinearity. With **314,335 bp** the MT325 genome is slightly smaller and does not exhibit long range colinearity with the PBCV-1 and NY-2A genomes. As in previously sequenced *phycodnaviridae*, new unexpected functions turned up to be encoded is these two new genomes, such as the first aquaglycerolporin the activity of which has been experimentally verified. A detailed comparative proteomics of the three viral particles (each of them exhibiting about 120 virus-encoded polypeptides) is also under way in the same laboratory (D. Dunigan *et al.*).

It was known for some time that filamentous marine brown alga of genus *Feldmannia* were infected by large dsDNA viruses (phaeovirus) coming in two genome sizes: 158kbp and 178kbp [3]. Prof. T-J. Choi, (Pukyong National University, Busan, Korea) reported on the completion of the genome sequencing of the "short" form of FsV infecting *Feldmannia sp*. The final sequence size is **153,259 bp** (51.8 G+C). About 50% of the 161 predicted ORFs have their best matching homologues in *Feldmannia irregularis* virus (FirrV-1) or *Ectocarpus Siliculosus* virus (EsV-1). Nicolas Delaroque (Max Plank Institute for Chemical Ecology, Jena, Germany) reported on the difficulty to reach full closure in sequencing the genome of *Feldmannia irregularis* virus (FirrV-1), most probably due to the presence of long repeats. The current FirrV genome sequence data consists of **191,667 bp** (in 16 contigs), encoding 156 putative proteins [4]. His more recent experiments suggest that FirrV infection may lead to the release of a mixture of virus forms associated with a wide range of genome sizes (from 192kbp to 10 kbp).

Dr Keizo Nagasaki and his collaborators (National Research Institute of Fisheries, Hiroshima, Japan) announced the near completion of the **356 kbp**-genome sequence of HcV01, a dsDNA virus infecting dinoflagelate *Heterocapsa circularisquama*, and of the **294 kbp**-genome sequence of *Heterosigma akashivo* infecting virus (HaV01). The same laboratory is also finishing the sequencing of the **145 kb**-genome of T4-like looking cyanophage Ma-LMM01, infecting the toxic cyanobacterium *Microcystis aeruginosa*.

Finally, Willie Wilson's group (Plymouth Marine Laboratory, UK) claimed the bronze medal in the fierce competition for genome size [5]. They presented the complete genome sequence of Coccolithovirus EhV-86, a ds-DNA infecting alien-looking calcarous nanoplankton ***Emiliania huxleyi*** (Fig.1). The genome is made of **407,339** bp (40.2% G+C) and encodes 472 putative protein coding regions. Only 66 (14%) of them have recognizable homologues in the public databases. As other giant viruses, EhV-86 exhibits its share of unexpected genes and functions, most notably a number of enzymes involved in the biosynthesis of sphingolipids. Albeit phylogenetically branching at the root of the *Phycodnaviridae* (e.g. PBCV-1 or EsV), EhV-86 does encode it own DNA-dependent RNA polymerase complex, thus filling the gap with the other Nucleo-Cytoplasmic Large DNA virus families (Irido-, Asfar-, Pox-, and Mimi-viridae) that all exhibit virally-encoded RNA-polymerases. Pending approval by ICTV, EhV-86 might become the prototype of the coccolythovirinae, a new subfamily of phycodnaviridae.

In his closing lecture, Jim Van Etten, reminded the new comers in the field of algal viruses that reports of very large icosahedral virus-like particles in various aquatic and marine organisms can be traced back to the 50's, but failed to elicit much interest outside of community of marine biologists. The discovery and genome characterization of the large freshwater chlorella viruses [2], and more recently of giant amoeba infecting Mimivirus [6] (remotely related to phycodnaviruses but not an algal virus) elicited a renewed interest in the genomics of these large marine viruses, as they may provide new insight on the early evolution of eukaryotes. Not unexpectedly, close relatives of Mimivirus appear to exist in the marine environment, as suggested by the numerous homologous sequences found by J.-M. Claverie and E. Ghedin (The Institute for Genomic Research, Rockville, USA) in their exhaustive analysis [7] of the Sargasso Sea environmental data set [8].

The 4th Algal Virus Workshop made it clear that these giant algal viruses are now entering the genomic era at full speed. The amount of surprises that we can expect while deciphering their genomes will be as big as their diversity, and more dogma on what a virus should look like will probably be shattered along the way.

**Acknowledgements**

Thanks to the participants and organizers of the 4[th] algal virus workshop (www.avw4.org/) for allowing some of their unpublished work to be mentioned in this article.

**Figure.1.** Scanning Electron Microscopy picture of *Emiliania huxleyi* [9]. Alien looking *E. huxleyi* is the host of phycodnavirus EhV-86, the 407-kb genome of which was sequenced at the Sanger center [10].

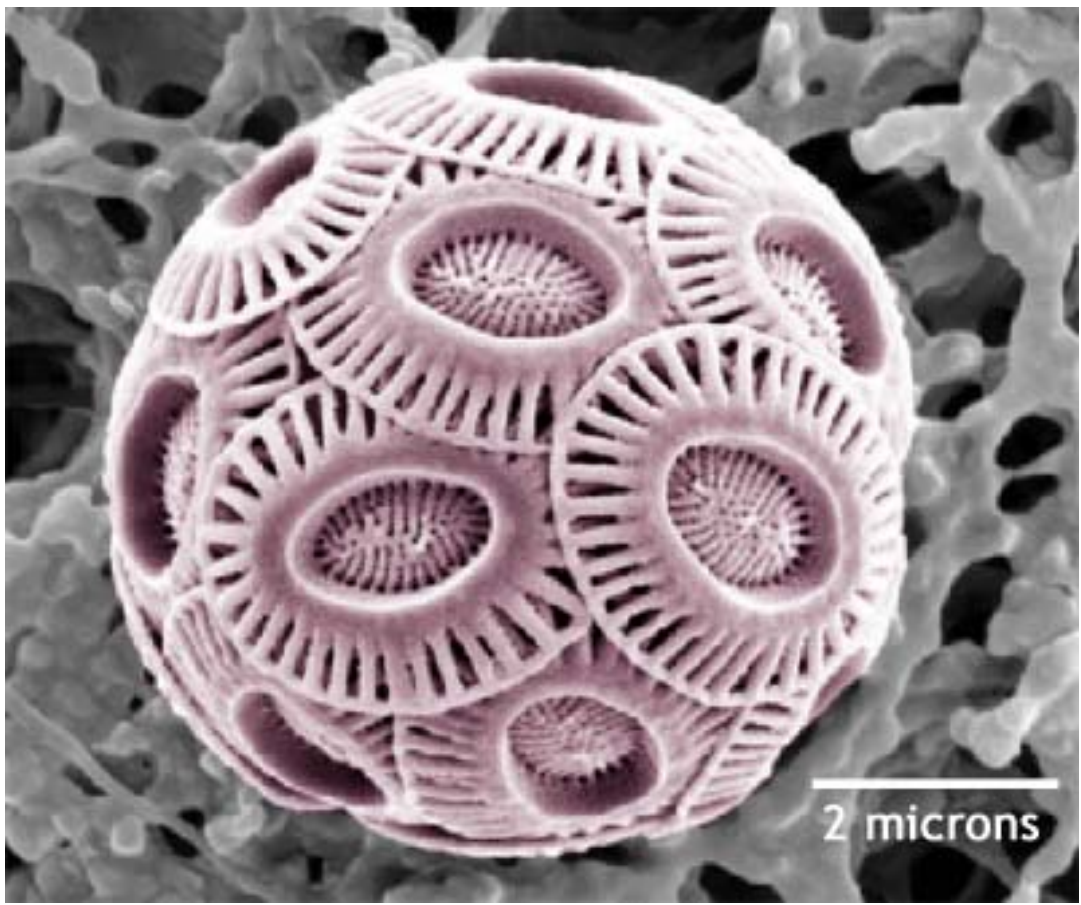